\documentclass[fleqn,10pt,twocolumn]{wlscirep}
\usepackage[utf8]{inputenc}
\usepackage[T1]{fontenc}
\usepackage{orcidlink}
\usepackage{graphicx}
\usepackage{dcolumn}
\usepackage{bm}
\usepackage{hyperref}
\usepackage{verbatim}
\usepackage[capitalise]{cleveref}
\usepackage[skins,theorems]{tcolorbox}
\usepackage{multirow}
\tcbset{highlight math style={enhanced,colframe=red,colback=white,arc=0pt,boxrule=1pt}}

\usepackage{xcolor}
\usepackage[normalem]{ulem}
\usepackage{bm}

\usepackage{textcomp}
\def\qe{\textsc{Quantum ESPRESSO}\texttrademark}

\def\sportran{\texttt{SporTran}}

\usepackage{textcomp} \def\qe{\textsc{Quantum ESPRESSO}\texttrademark}

\definecolor{tangerine}{rgb}{0.944,0.522,0}
\definecolor{verde}{rgb}{0.,0.6,0}
\definecolor{rosso}{rgb}{0.9,0.0,0.2}
\definecolor{arancio}{rgb}{0.9,0.6,0.4}
\definecolor{viola}{rgb}{0.9,0.3,0.9}
\definecolor{turchese}{rgb}{0,0.5,0.5}

\newcommand{\editor}[2]{%
  \expandafter\newcommand\csname #1note\endcsname[1]{%
    \textcolor{#2}{(\textbf{#1:} ##1)}}%
  \expandafter\newcommand\csname #1\endcsname[1]{%
    \textcolor{#2}{##1}}%
  \expandafter\newcommand\csname #1cancel\endcsname[1]{%
    \textcolor{#2}{\sout{##1}}}%
  \expandafter\newcommand\csname #1change\endcsname[2]{%
    \textcolor{#2}{\sout{##1} ##2}}%
  \newenvironment{#1text}{\color{#2}}{\color{black}}
}

\editor{SB}{tangerine}
\editor{DT}{blue}
\editor{CM}{red}
\editor{RC}{viola}
\editor{LZ}{verde}

\title{Viscosity in water from first-principles and deep-neural-network simulations}

\newcommand{\ai}{\textit{ab initio} }

\author[1]{Cesare Malosso\orcidlink{0000-0002-2760-3632}}

\author[2,3]{Linfeng Zhang}

\author[2,4]{Roberto Car}

\author[1,5]{Stefano Baroni\,\orcidlink{0000-0002-3508-6663}}

\author[1]{Davide Tisi\,\orcidlink{0000-0001-7229-6101}}

\affil[1]{
    SISSA -- Scuola Internazionale Superiore di Studi Avanzati, 34136 Trieste, Italy
}
\affil[2]{
Program in Applied and Computational Mathematics, Princeton University, Princeton, NJ 08544, USA
}
\affil[3]{
DP Technology, Beijing 100080, People’s Republic of China
}
\affil[4]{
    Department of Chemistry, Department of Physics, and Princeton Institute for the Science and Technology of Materials, Princeton University, Princeton, NJ 08544, USA
}
\affil[5]{
    CNR Istituto Officina dei Materiali, SISSA unit, 34136 Trieste, Italy
}
\affil[*]{baroni@sissa.it}

\date{\today}

\begin{abstract}
    We report on an extensive study of the viscosity of liquid water at near-ambient conditions, performed within the Green-Kubo theory of linear response and equilibrium \emph{ab initio} molecular dynamics (AIMD), based on density-functional theory (DFT). In order to cope with the long simulation times necessary to achieve an acceptable statistical accuracy, our \emph{ab initio approach} is enhanced with deep-neural-network potentials (NNP). This approach is first validated against AIMD results, obtained by using the Perdew-Burke-Ernzerhof (PBE) exchange-correlation functional and paying careful attention to crucial, yet often overlooked, aspects of the statistical data analysis. Then, we train a second NNP to a dataset generated from the Strongly Constrained and Appropriately Normed (SCAN) functional. Once the error resulting from the imperfect prediction of the melting line is offset by referring the simulated temperature to the theoretical melting one, our SCAN predictions of the shear viscosity of water are in very good agreement with experiments. 
\end{abstract}
\begin{document}

\flushbottom
\maketitle
\thispagestyle{empty}

%\noindent Please note: Abbreviations should be introduced at the first mention in the main text – no abbreviations lists. Suggested structure of main text (not enforced) is provided below.

\section{\label{sec:I}Introduction}
Shear viscosity is one of the most important transport properties governing the macroscopic flow of liquids. As such, it plays a fundamental role in various fields of science and technology, such as, \emph{e.g.}, chemical and mechanical engineering or earth and planetary sciences, to name but a few. For instance, the viscosity of a solvent crucially affects the dynamics of solutes and the reactions rates, of fundamental importance in the study of biological processes and chemical reactions \cite{chem2,chem1,chem3}. The value of the viscosity of liquid iron, abundant in Earth's outer core, is key in the prediction of the magnetic field of rocky planets  \cite{Wijs1998,Alfe1998}. An accurate determination of the temperature and pressure profile of the viscosity is also essential for the correct modelling of tidal interactions in the planets' interior, in particular in the presence of icy layers \cite{tidalFriction2020,Dumoulin2017}.

In this work we focus on water, an ubiquitous molecular liquid with extraordinary and complex properties \cite{waterChemRev,Pourasad2021,Lu2008,Sharma2005,Sharma2007,Gartner26040}. In spite of the great importance of this system and the large number of studies based on density-functional theory (DFT) and \emph{ab initio} molecular dynamics (AIMD) devoted to it \cite{Parrinello2004,Grossman2004,Schwegler2004,Teodora2006,Gillan2016,Chen10846,gygi,Zheng2018SCAN,kuhne},
all of these efforts have, until very recently \cite{Joly}, dodged its viscous properties, because \emph{an accurate computation of the viscosity of water would require exceedingly long first-principles simulations} \cite{gygi}. A number of studies based on classical force fields exists \cite{abascal,Tazi,Heyes2019,MonterodeHijes2018}, but the poor transferability of these models sets a limit to their predictive power.  An attempt to estimate the viscosity of water from first principles was made with an indirect approach relying on the Stokes-Einstein relation \cite{kuhne}, which, however, does not hold over all the phase diagram for liquid water, particularly in the supercooled regime \cite{Kumar9575,Xu2009,Tsimpanogiannis}. 

A rigorous microscopic description of the shear viscosity of liquids, $\eta$, is provided by the Green-Kubo (GK) theory of linear response \cite{Green,Kubo,Evans2007,Allen2017}, according to which its value is proportional to the integral of the time auto-correlation function (tACF) of the off-diagonal matrix elements of the stress tensor. This integral can be estimated from the time series of the stress, generated by an equilibrium molecular-dynamics simulation of the system of interest. A number of different procedures have been developed to cope with the evaluation of the GK integral \cite{Maginn_Messerly_Carlson_Roe_Elliot_2018,Maginn2}. Here, we adopt a spectral approach, recently proposed by Ercole \emph{et al.}  \cite{Ercole2017,Ercole2016,Baroni2018,Grasselli2021}, which allows one to compute transport coefficients, along with the statistical errors affecting them, from shorter trajectories than previously thought to be necessary. This progress notwithstanding, the estimate of transport coefficients from AIMD may require generating trajectories of a few hundred picoseconds for systems as large as a few hundred atoms. It is evident that, although technically quite possible, AIMD simulations of this size do not lend themselves to an easy estimate of the statistical accuracy of the results, let alone a systematic exploration of a broad region of the phase diagram of a material. 

The last decade has seen the rise of machine-trained potentials, as represented by either deep-neural networks \cite{Behler2007,Smith2017,Linfeng2018,RevBehler} or by Gaussian-processes \cite{Bartok2010}, as powerful tools for atomistic simulations. These potentials are able to deliver a nearly quantum mechanical accuracy at a cost that is only marginally higher than that of classical force fields. This opens the way to extend the scope of AIMD simulations to the size range necessary for the computation of reliable transport coefficient such as the viscosity.
In the present work we adopt the recently developed \emph{Deep Potential} framework \cite{Linfeng2018,NIPS2018_7696,100_Milion_Atoms} to study the shear viscosity of liquid water. Deep potential molecular dynamics (DPMD) simulations have already been proved to successfully predict bulk thermodynamic properties beyond the reach of direct DFT calculations \cite{DeepWater2021,Jiang2021,zhang2021modeling,Wu2021,Gartner26040,Niu2020}, as well as dynamic properties like mass diffusion in solid state electrolytes \cite{Marcolongo2019,huang2021deep}, their interactions with defects\cite{pegolo2021temperature}, thermal transport properties in silicon \cite{LIDeepMD}, infrared spectra of water and ice \cite{DeepWannier}, Raman spectra of water \cite{Sommers2020} and very recently also the thermal conductivity of liquids such as liquid water \cite{dtisi}. 

So far, a combination of AIMD, advanced data analysis, and neural-network techniques has only been applied to thermal and charge transport \cite{Baroni2018,Grasselli2020,Grasselli2021,pegolo2021temperature,Marcolongo2021,dtisi}. In this work we attempt to apply them to the computation of viscosity. In this study we report on calculations, from both direct DFT and DPMD simulations, of the shear viscosity of water. We show that $\eta$ can be obtained with trajectories of $\approx 400 $ ps, that are still quite demanding for a extensive \ai study over a broad portion of the phase diagram. We thus take advantage of the DPMD technique and perform extensive simulation employing a deep-neural-network potential (NNP) trained on extensive DFT data. Our methodology proceeds in two steps. In the first, we train a NNP on PBE \cite{PBE} data and validate our procedure against results from a rather long ($400$-ps) AIMD trajectory. We then adopt the  \emph{strongly constrained and appropriately normed} (SCAN) meta-GGA exchange-correlation (XC) functional \cite{Sun2016SCAN,SCANPerdew}, which provides a much more accurate description of the H-bond network in water \cite{Chen10846}, to perform extensive simulations of the viscous properties of water just above melting. Close to melting, the viscosity depends very sensitively on temperature. Once the error resulting from the imperfect prediction of the melting line is offset by referring the simulated temperature to the theoretical melting one, our SCAN predictions of the shear viscosity of water in a temperature range extending above the melting line are in very good agreement with experiment.

Our paper is organized as follows. \cref{sec:results} contains all the discussion of the results: in \cref{sec:AIMD} we present the results of our direct PBE-AIMD simulations and draw some conclusions on the simulation time and length scales necessary to achieve an acceptable statistical accuracy; in \cref{sec:PBE-NNP} we benchmark our NNP against \emph{ab initio} MD simulations of liquid water at the PBE level of theory; in \cref{sec:PBE-analysis} we expand our analysis on the statistical properties of our estimator of the shear viscosity and briefly discuss its size-dependency. 
Once our methodology is set up and validated, in \cref{sec:SCAN} we report on an extensive set of simulations performed with a NNP model trained on SCAN meta-GGA DFT data and
we compare their results with available experimental data and our PBE-NNP results. We show that SCAN meta-GGA reduces the deviation from experiments of the predicted shear viscosity. Section \ref{sec:conclusion} contains our final discussion with some interesting perspective and further applications of our work. In \cref{sec:II}, we recall the main theoretical and numerical methods used throughout the work: the main aspects of the GK theory of transport; its application to viscosity; the main data-analysis technique; and briefly describe the neural-network model.

\section{Results}\label{sec:results}

\subsection{\label{sec:AIMD}{Ab initio Molecular Dynamics}}

\begin{figure}[t!]
    \centering
    \includegraphics[width=7.5cm]{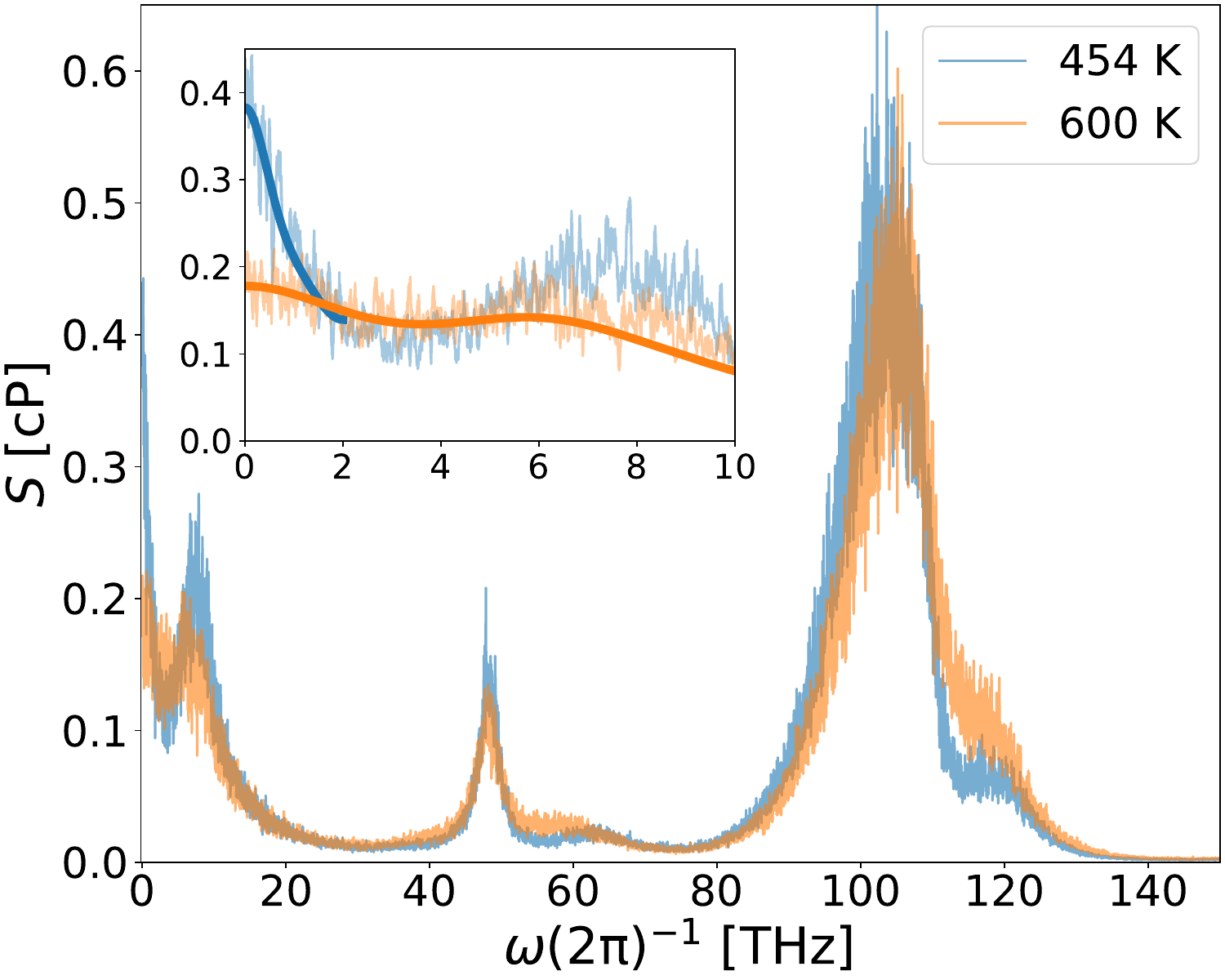}
    \caption{\textbf{\textit{Ab initio} power spectra.} Power spectra of the off-diagonal elements of the stress in water at $454$ K (blue) and $600$ K (orange), obtained from  AIMD simulations (see text). The spectra are filtered by a moving average with a window of $0.05$ THz. The thick solid lines in the inset represent the cepstral-filtered spectra whose zero-frequency value gives an estimate of the shear viscosity.}
    \label{fig:abinitio_psd}
\end{figure}

We performed AIMD simulations of liquid water at near-ambient conditions using the Perdew-Burke-Ernzerhof (PBE) \cite{PBE} XC 
functional, the plane-wave pseudopotential method, Hamann-Schluter-Chiang-Vanderbilt (HSCV) norm-conserving pseudopotentials \cite{Hamann2013}, and a kinetic-energy cutoff of $85$ Ry. The simulated system was made of $64$ molecules at the standard density of 1 gr\,cm$^{-3}$, corresponding to a cubic box of edge $l=12.43$ \AA. All the simulations were carried out with the Car-Parrinello extended-Langrangian method \cite{Car:1985} using the \texttt{cp.x} component of the \qe\ distribution \cite{Giannozzi_2009,Giannozzi_2017,QE3} and setting the fictitious electronic mass to $25$ physical masses and the timestep to $dt = 0.073$ fs. We performed two simulations aiming at thermodynamic conditions near ambient temperature and somewhat
above it. As PBE is known to enhance the short-range structure of water and to overestimate the melting temperature by $\approx 140$ K \cite{Sit2005,Yoo2009}, we set the target temperatures of the two simulations to $450$ and $600$ K, respectively. Both trajectories where first equilibrated in the NVT ensemble using a Nos\`e-Hoover thermostat \cite{nosehoover} at the target temperature, followed by long production NVE runs of $400$-ps. Finally, the shear viscosity
was obtained from the cepstral analysis of the power spectrum of the off-diagonal elements of the stress, using the \sportran\ \cite{SporTran} code.

\begin{figure}[t]
    \centering
    \includegraphics[width=7.5cm]{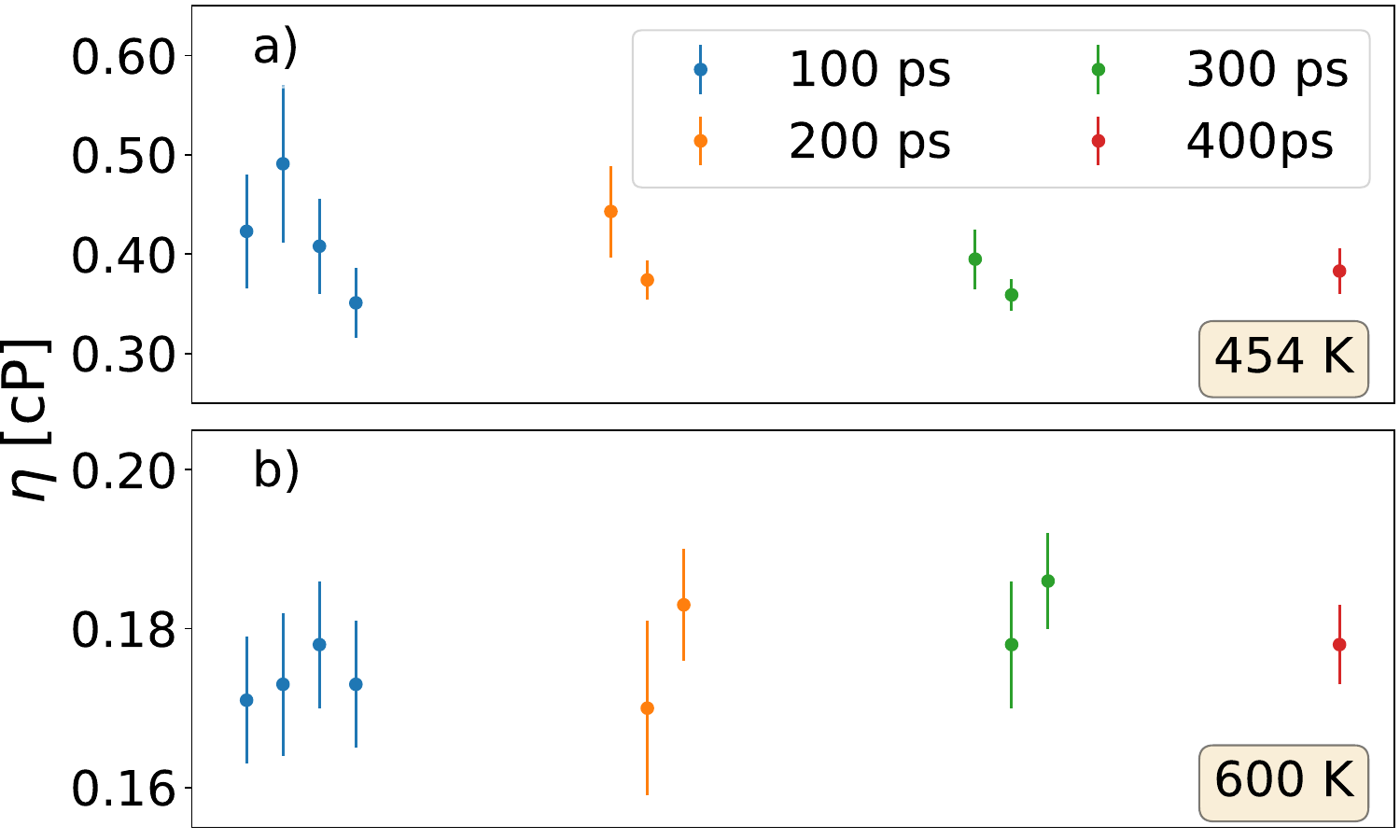}
    \caption{\textbf{$\eta$ vs. length of the simulation.} Dependency of the shear viscosity $\eta$ on the length of the simulation, estimated by AIMD (a) at $454$ K and (b) $600$ K . Different colors refer to different simulation times. Error bars represent standard deviations.}
    \label{fig:convergency_ai}
\end{figure}

In Fig. 1 we display the (moving averages \cite{MovingAverage} of the) power spectra of the stress-tensor time series resulting from our two simulations. While showing similar features at high frequency, the two spectra differ substantially approaching $\omega=0$. In particular, lower temperatures see the appearance of sharp peaks near $\omega=0$, which requires a greater care in the cepstral analysis of the data, which is based on a low-pass filter of the (logarithm of) the power spectra. In the inset we display the low-frequency region of the spectra together with the results carried out by the cepstral analysis, \emph{i.e.} by applying a low-pass filter to the logarithm of the raw spectra. The filtered spectra are represented by thick solid lines whose zero-frequency value is a fair and accurate estimate of the shear viscosity we are after:
\begin{equation*}
    \label{eq:results_ai}
    \eta = \left \{ 
    \begin{matrix}
        ~ 0.383 \pm 0.023 \ \text{cP} \quad \text{at 454 K,}\\
        ~ 0.178 \pm 0.005 \ \text{cP} \quad \text{at 600 K.}
    \end{matrix}
    \right .
    \quad\quad\text{(PBE)}
\end{equation*}
where the unit cP stays for \emph{centipoise},  $1 \ \text{cP}=10^{-3} \ \text{Pa} \cdot \text{s} $. It is often assumed that the predictions of \emph{ab initio} simulations should be compared with experiment upon shifting the simulated 
temperature by the offset between the theoretical and experimental melting temperatures, which, in the case of PBE, amounts to $T_\mathrm{m} (\mathrm{PBE}) -T_\mathrm{m}(\mathrm{expt}) \approx 140~\mathrm K$ \cite{Yoo2009}. We thus compare our value predicted by PBE at $T=454~\mathrm K$ with the experimental value measured at $T=313 \approx  454-140~\mathrm K$, $\eta^{\mathrm{expt}}(T=313~\mathrm K)=0.653~\mathrm {cP}$.
The agreement is fair, on account of both the uncertainties related to the empirical temperature shift and the very sensitive dependence of the viscosity upon temperature near melting. More on the meaning of the residual disagreement will be discussed in Sec. \ref{sec:SCAN}.

\begin{figure}[t]
   \centering
   \includegraphics[width=7.5cm]{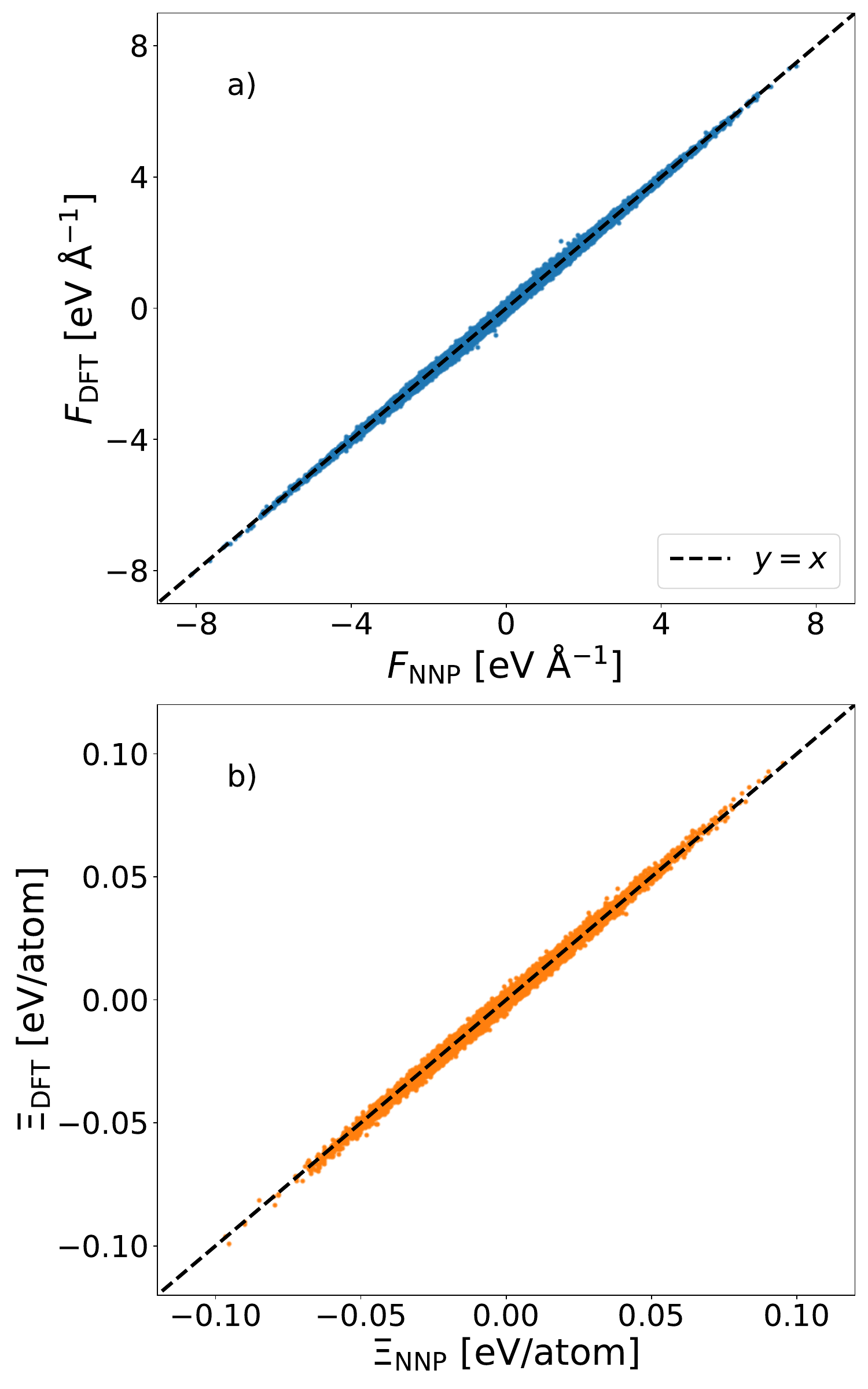}
   \caption{\textbf{Neural network accuracy.} Scatter plot of the NNP forces (a) and off-diagonal elements of the virial per atom (b) \emph{vs.} DFT data, for a dataset of $10000$ configurations. The corresponding correlation coefficients are  $R_a^2=0.998$  and $R_b^2=0.995$.}
   \label{fig:accuracy}
\end{figure}

In Fig. 2 we display how the prediction of the shear viscosity in water depends on the length of the simulation. In order to highlight the impact of possibly long relaxation times on the estimate of the transport coefficient, we have split our $400$-ps trajectories into segments of 100, 200, and 300-ps (in the latter case the two segments were overlapping).The estimates from different segments coincide within the statistical errors evaluated within each of them at 600K, but not quite so at 454 K. This can be ascribed to the emergence of a narrow peak in the stress power spectrum at $\omega =0$ (see Fig. 1), related to an increase of the stress correlation time occurring as the freezing temperature is approached. A similar behaviour had been already observed by Ercole \emph{et al.} \cite{Ercole2017} in the case of heat transport in strongly harmonic crystals. All these considerations suggest that near freezing the computation of the shear viscosity requires longer simulation runs, and even longer runs would be required for a fair evaluation of the statistical uncertainties, indicating that AIMD may not be the most efficient approach to explore a broad range of thermodynamic conditions. In the following we show that neural-network models of inter-atomic interactions trained on \emph{ab initio} data provide a valid alternative to direct AIMD simulations, yielding results of similar quality at a much lower computational cost.

\subsection{PBE NNP}\label{sec:PBE-NNP}
\begin{figure}[t]
    \centering
    \includegraphics[width=7.5cm]{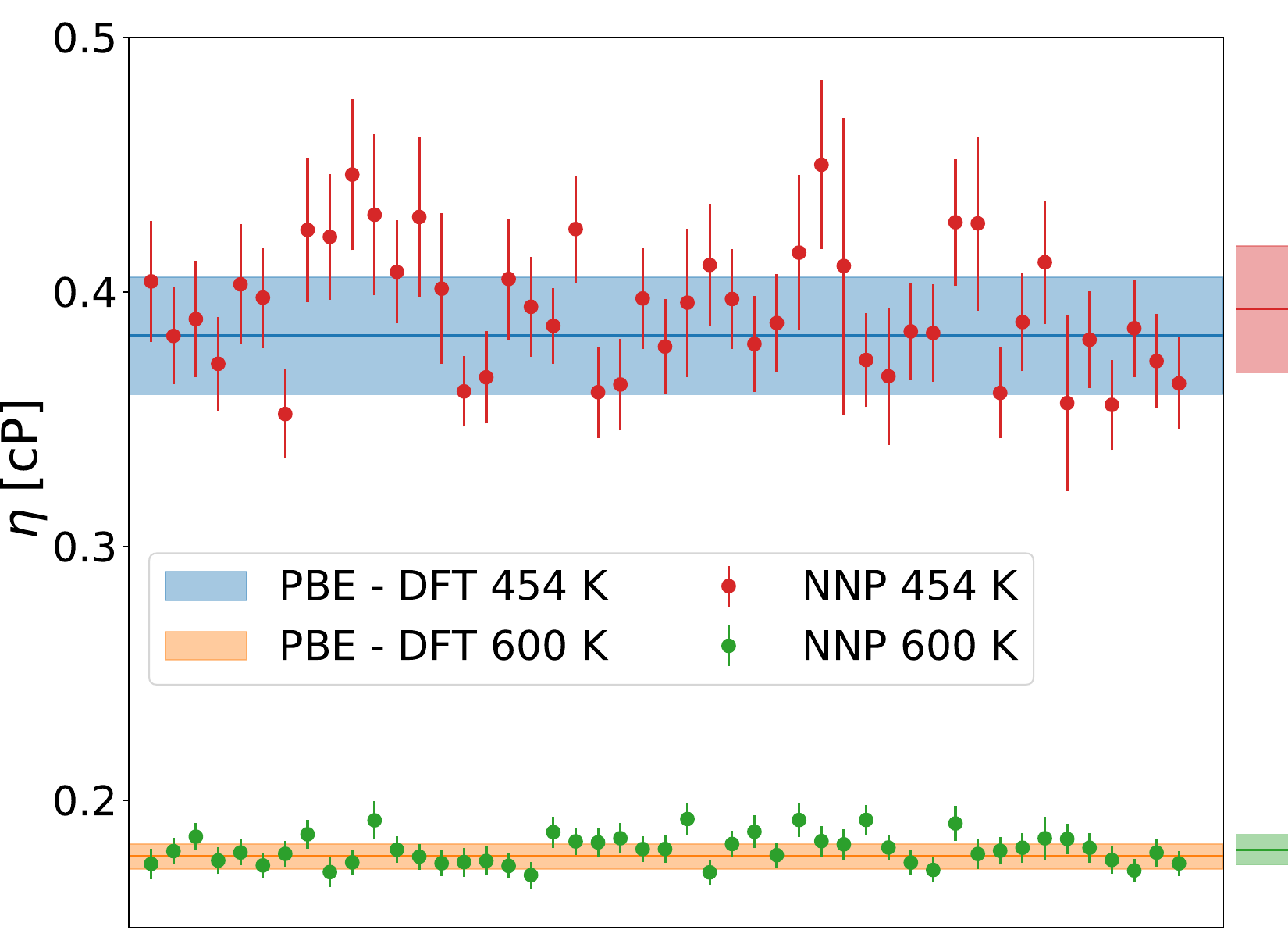}
    \caption{\textbf{Comparison between the shear viscosity predicted by DFT AIMD simulations and by DPMD simulations.} The results are obtained from $400$-ps long DFT AIMD simulations (horizontal blue and orange bands, 454 and 600 K respectively) and by DPMD simulations of the same length (solid dots). The width of the bands and the vertical bars across the dots indicate the standard deviation of the data they refer to, as estimated by cepstral analysis (see Sec. \ref{sec:IIa}). The red and green bars on the right of the box indicate the sample averages and standard deviations of the DPMD data. Error bars represent standard deviations.}
    \label{fig:benchmark}
\end{figure}

\begin{figure*}[ht]
    \centering
    \includegraphics[width=7.5cm]{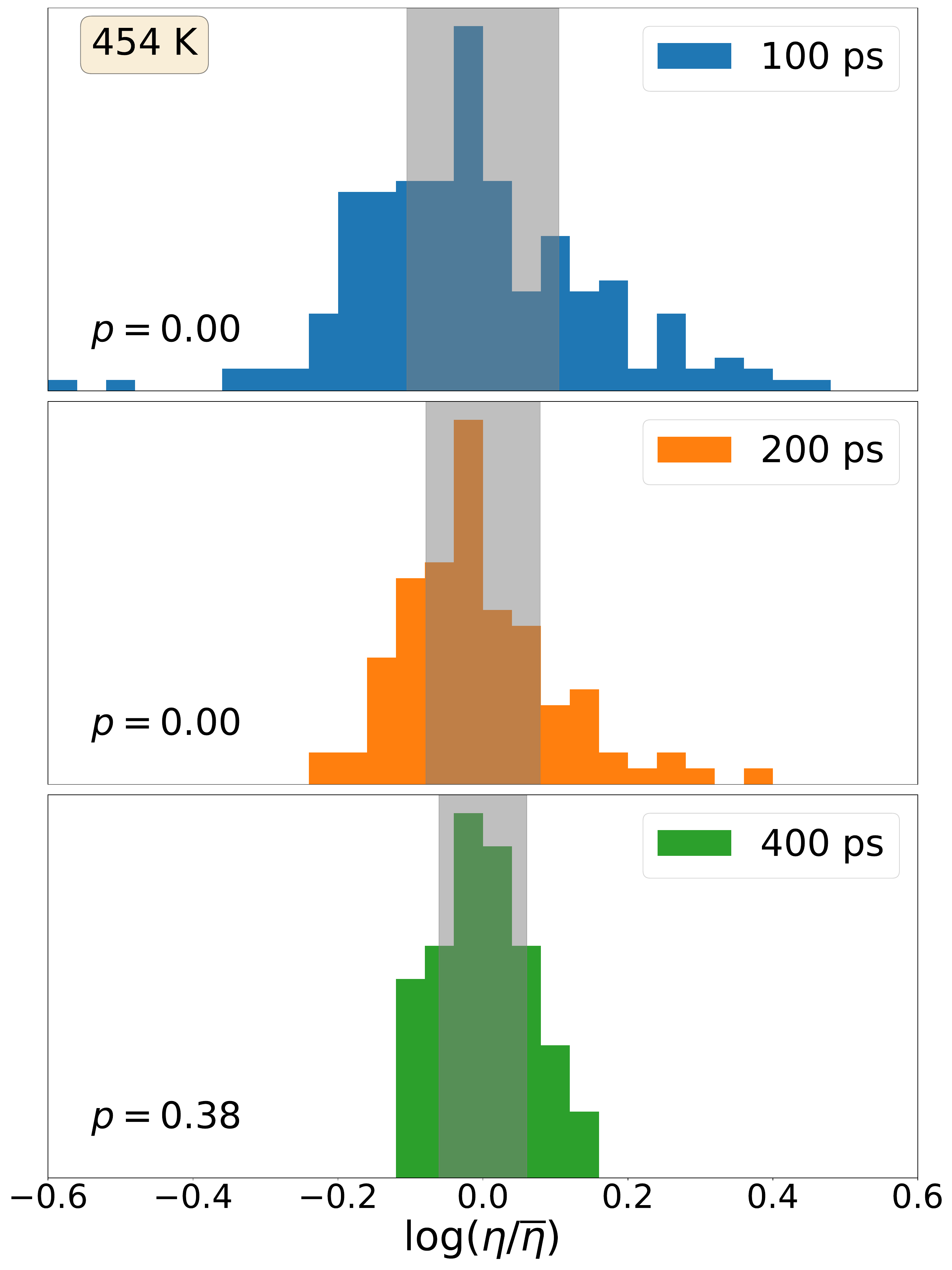}
    \quad \includegraphics[width=7.5cm]{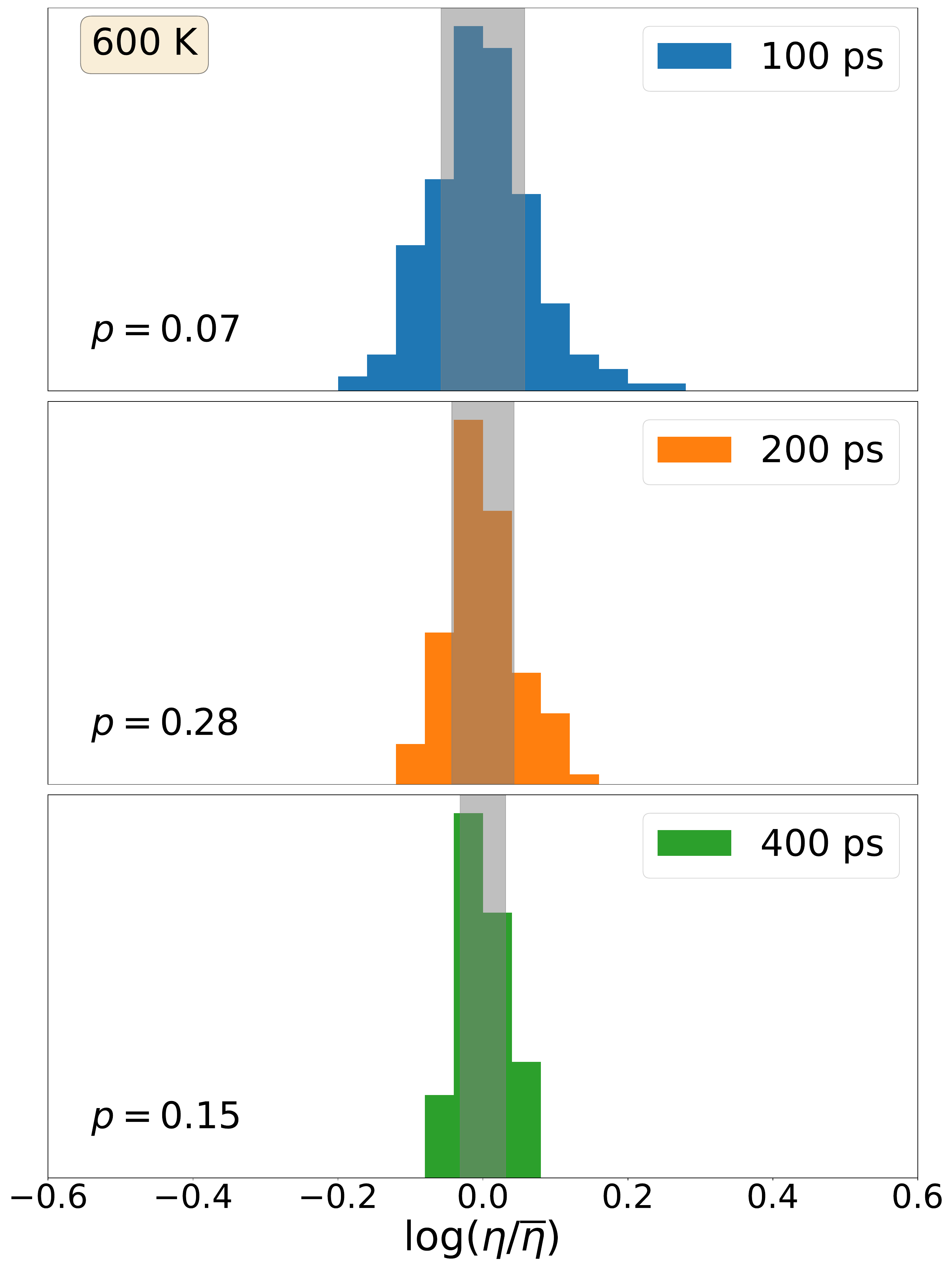}
    \caption{\textbf{Normalized distributions of the logarithm of the shear viscosities, log($\eta$).} The results are estimated over multiple MD segments (blue: $100$-ps; orange: $200$-ps; green: $400$-ps) extracted from a $20$-ns trajectory at $454$ K (left) and $600$ K (right). The reported data are referred to  the average, $\bar{\eta}$. We remind that the absolute error on log($\eta$) is the relative error on $\eta$. The shaded area denote the average standard deviation of the shear viscosities, as estimated by cepstral analysis within each individual segment.}
    \label{fig:histo}
\end{figure*}

In order to appraise the ability of NNP to accurately predict shear viscosity, we have generated one such model, by training it on a set of PBE-DFT data. The training dataset is prepared via a recently proposed ``on-the-fly'' learning procedure called Deep Potential Generator (DP-GEN) \cite{DPGEN,Linfeng2019_dpgen} and it consists of the energies and atomic forces of $4000$ configurations of water generated by the DP-GEN from NPT MD trajectories at different temperatures in the [$300$-$700$ K] range and for pressures up to $50$ kbar. The PBE-NNP is then constructed and trained with the DeePMD-kit. The cutoff radius is set to $6$ Å. The size of the embedding and fitting nets is ($50$, $50$, $50$) and ($250$, $250$, $250$), respectively. The model was trained by minimizing the standard loss function, $\mathcal{L}$, presented in Equation \eqref{eq:loss-function} of \cref{sec:II} with 2 million steps of Adam stochastic gradient descent \cite{Kingma2014}. We tried to include the values of the virial in the definition of the loss function, but we found no improvement with respect to the standard definition of Equation \eqref{eq:loss-function}, and thus decided not to modify it.

Fig. 3 shows a scatter plot of the NNP predictions for atomic forces and stress \emph{vs.} PBE-DFT data, evaluated over a set of 10000 configurations, not included in the training dataset. The average error on the forces and on the off-diagonal elements of the virial are  $\sigma_F = 40 $ meV\,\AA$^{-1}$ \ and $\sigma_{\Xi} = 1.4$ meV/atom, respectively, corresponding to correlation coefficients \cite{CorrCoeff} of 0.998 and 0.995, respectively.

\begin{figure}[ht]
    \centering
\includegraphics[width=7.5cm]{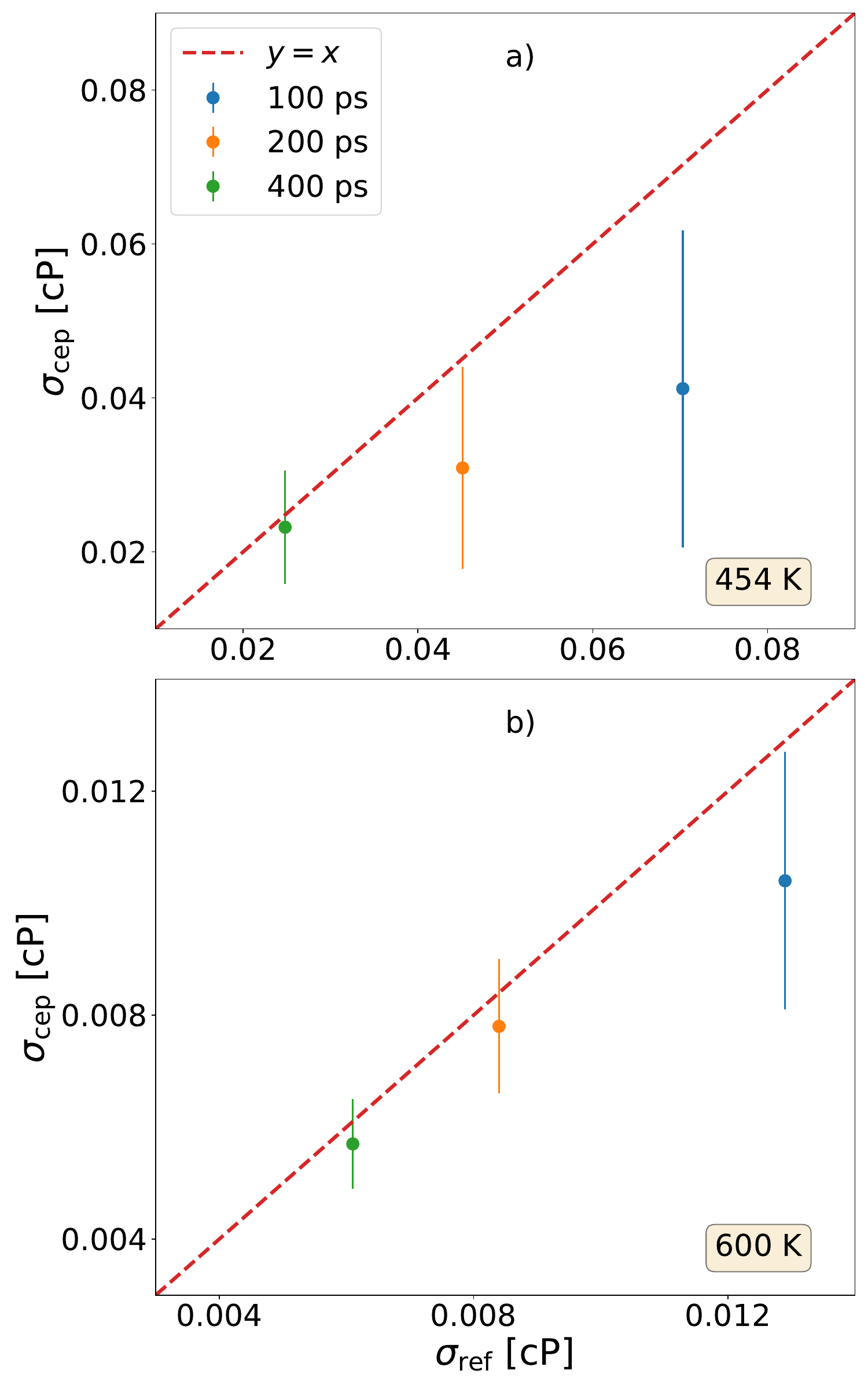}
    \caption{\textbf{Estimated standard deviations of $\eta$ vs. spread of the distribution.} Correlation between the cepstral estimates of the standard deviations of the viscosity of water from trajectories of different lengths and temperatures, $\sigma_{\text{cep}}$, \emph{vs.} the spread of the distribution of their values resulting from different trajectories, $\sigma_{\text{ref}}$ (see text). Error bars represent standard deviations.
    }
    \label{fig:sigma}
\end{figure}

\begin{figure}[ht]
    \centering
    \includegraphics[width=7.5cm]{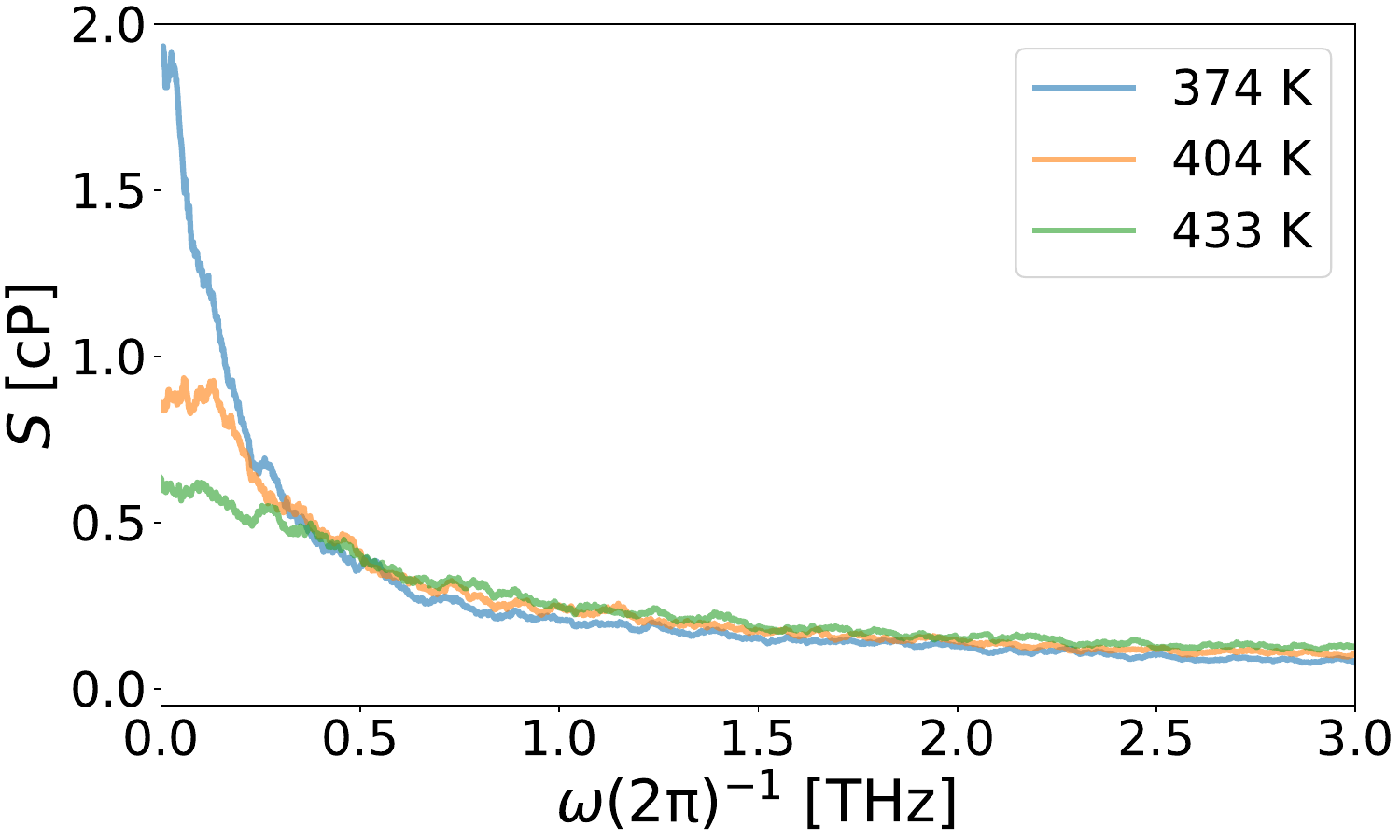}    
    \caption{\textbf{Moving average of the low-frequency region of the power spectrum of the off-diagonal elements of the stress tensor in water at different temperatures, as obtained from DPMD simulations trained on PBE DFT data.} An averaging window of $0.05$ THz was used. Simulations were run at the fixed density of 1 gr\,cm$^{-3}$.}
    \label{fig:psd_low}
\end{figure}

In order to validate our neural-network methodology for the prediction of the shear viscosity, we performed DPMD simulations in the \emph{NVE} ensemble for the same model of liquid water described above. Simulations of $20$-ns were run at two different temperatures, $454$ K and $600$ K, using our NNP trained to PBE water. All simulations were carried out using the LAMMPS code \cite{Thompson2021} interfaced with DeepMD-kit. In Figure 4 we display the results obtained by analysing independently each one of the 50 $400$-ps segments in which we have partitioned the whole $20$-ns trajectory. The shear viscosity of each segment is obtained again by cepstral analysis using the \sportran\ code and is represented by solid dots together with its estimated statistical error. The blue and orange regions represent respectively the estimate of the shear viscosity given in \cref{sec:AIMD} from \emph{ab initio} MD simulations at $454$ K and $600$ K. We observe a very good agreement between the two  approaches and conclude therefore that our NNP is capable of predicting correctly the shear viscosity of water at the given pT conditions. Also notice the close agreement between the standard deviation of the viscosity estimated by cepstral analysis on individual 400-ps trajectory segments and the value computed over a sample of 50 segments. More on the statistical analysis and significance of our data in Sec. \ref{sec:PBE-analysis}.

In Table 1 we report our results for the viscosity of water computed at two different temperatures with DPMD and NNP trained on PBE-DFT data, obtained from very long ($20$-ns) trajectories, and compare them with the AIMD data of Sec. \ref{sec:AIMD}.

\begin{table}[ht]
\caption{\label{tab:tableresult}
Viscosity of water [cP] computed from AIMD or DPMD performed at two different temperatures using the PBE XC functional.
} 
\centering
\begin{tabular}{lcc}
\textrm{}&
\textrm{T $= 454$ K}&
\textrm{T $= 600$ K} \\
\hline

\text{AIMD} &  0.383 $\pm$ 0.023 & 0.178 $\pm$ 0.005 \\
\text{DPMD}  &  0.402 $\pm$ 0.005 & 0.184 $\pm$ 0.001  \\

\end{tabular}
\end{table}

\subsection{Statistical analysis and finite-size scaling}\label{sec:PBE-analysis}
We are now ready to investigate the statistical behaviour of the shear viscosity for different simulation lengths. To this end, we sliced our $20$-ns simulations in segments of smaller lengths ($100$-, $200$-, and $400$-ps) and analyzed them independently. Before proceeding, we remind the pivotal tenet of cepstral analysis: if a sample of a stationary stochastic process is longer than all the relevant time scales of the process, then the sample spectrum (\emph{i.e.} the squared modulus of the Fourier transform of the series) equals the theoretical power spectrum of the process, times a set of identically distributed $\chi^2$ stochastic variables that are independent from each other for different frequencies. This implies that the low-pass-filtered logarithm of the sample spectrum is normally distributed at any (sufficiently low) frequency \cite{Ercole2017} and that the estimator of the transport coefficient---which is proportional to the $\omega=0$ value of the filtered spectrum---is therefore a log-normal variate. In order to check the reliability of the cepstral estimate of the viscosity from trajectories of different lengths, in Fig. 5 we display the distributions of the logarithm of these estimates from trajectory segments of different length ($100$-, $200$-, and $400$-ps) and report the p-values of the Shapiro-Wilk (SW) normality test \cite{SHAPIRO1965} for each distribution. We observe that: \emph{i)} at $T\approx 450~\mathrm K$ the WS test is failed for segments shorter than 400 ps, indicating the subsistence of slow stress fluctuations that adversely affect our data analysis technique ; \emph{ii)} at $T=600~\mathrm K$ the WS is never failed with respect to a standard significance level $\alpha=0.05$; even for the shortest segment length (100 ps), for which we compute a p-value of 0.07 over a sample of 200 segments, we have found that the distributions of $\log(\eta)$ resulting from smaller samples never fail the WS test with respect to this significance level; \emph{iii)} the width of the distributions of the viscosity estimated at different lengths is slightly larger than the standard deviation estimated within each segment by cepstral analysis; \emph{iv)} this difference decreases as the length of the segments increases, until it roughly vanishes at $400$-ps; \emph{v)} this difference also decreases by increasing the temperature. This observation is made more quantitative in Fig. 6, which shows the correlation between the standard deviations of the cepstral estimates of the viscosity from trajectories of different lengths and temperatures, \emph{vs.} the spread of the distribution of their values resulting from different trajectories. The former quantity is itself affected by a statistical uncertainty because cepstral analysis returns different standard deviations for different trajectories of a same length. Fig. 6 indicates that as the system approaches freezing from above and the viscosity increases, the low-frequency components of the virial fluctuations become increasingly important, and simulations of increasing length become necessary to cope with them. This is confirmed in Fig. 7 that displays the low-frequency portion of the power spectrum of the off-diagonal elements of the stress in water at different temperatures, and shows that as the system approaches freezing from above, a narrow peak develops at $\omega=0$, as a consequence of the onset of long-lived relaxation modes. In the present case, it appears that at $450$ K trajectories as long as $400$-ps are needed to get a reliable estimate of the statistical error affecting the estimate of the PBE-DFT viscosity. More generally, it seems that the flexibility offered by NNP and the long simulations they can afford are instrumental not only in exploring broad regions of the phase diagram of a material, but also in providing a reliable estimate of the statistical accuracy of individual simulations.

Finite-size effects may affect the transport properties calculated in numerical simulations \cite{Yeh2004,Grasselli2022}. In order to quantify these effects in the present case, we run up to 5-ns long NVE simulations at $454$ K and $600$ K of PBE-NNP water at fixed density and increasingly larger cells (with up to 4096 molecules). The results, reported in Table 2, indicate that $\eta$ shows no evident size dependence within the error bars of our simulations.

\begin{table}[ht]
\caption{\label{tab:tableSizeScaling}
Shear viscosity [cP] computed for water at different temperatures with PBE-NNP force field and using simulation boxes of different size. 
}
\centering
\begin{tabular}{lccc}
&
\multicolumn{3}{c}{size (number of molecules)}\\
\textrm{T [K]}&
64&
512&
4096 \\
\hline
 454  & 0.402 $\pm$ 0.005  & 0.402 $\pm$ 0.005 & 0.417 $\pm$ 0.007  \\
600 & 0.184 $\pm$ 0.001  &
        0.186 $\pm$ 0.002 &
        0.186 $\pm$ 0.002 \\
\end{tabular}
\end{table}

\subsection{SCAN NNP} \label{sec:SCAN}
The SCAN meta-GGA XC functional has demonstrated the ability to predict well several properties of water over a broad range of thermodynamic conditions, whose exploration was made possible by NNP techniques \cite{Chen10846,gygi,Gartner26040,DeepWater2021,Piaggi2021}. A combination of AIMD and NNP techniques, based on the SCAN XC functional, has recently been successfully applied to the prediction of the heat transport properties of liquid water \cite{dtisi}. In the following we report on our extension of this effort to the computation of the shear viscosity.

Accurate DPMD simulations were performed using NNP force fields trained on both PBE and SCAN DFT data \cite{dtisi} and the same software setup as in Sec. \ref{sec:PBE-NNP}. Our simulated systems consist of $512$ water molecules. With systems of this size, temperature fluctuations are smaller than 1K. We first perform NVT simulations at the target temperature, followed by NVE production runs, up to $5$-ns long. The volume was fixed to the value corresponding to the equilibrium densities evaluated in Ref. \cite{Piaggi2021} via enhanced-sampling simulations for SCAN, while for PBE it is computed from direct DPMD NPT simulations at ambient pressure, whose results are in agreement with previous calculations \cite{Gillan2016,gallih2o}.

In Fig. 8 we compare our SCAN-NNP and PBE-NNP results with each other and with experimental data \cite{haynes:2005,Dehaoui12020}. Results below the melting temperature, $T_m$, refer to the undercooled fluid, which becomes increasingly viscous as the temperature decreases. Remarkably, when temperatures are referred to the theoretical melting one, the SCAN predictions for the viscosity are in close agreement with experiment at melting (and above, as we will discuss shortly). This is not so for PBE. One could argue that PBE yields too low a viscosity as a consequence of the too low equilibrium density ($0.77$ \emph{vs} $\approx 1$ gr\,cm$^{-3}$ at melting). This is not the case, however, because repeating the simulations at the density of 1 gr\,cm$^{-3}$ (dashed lines) results in only a marginal increase in the predicted viscosity. We conclude that the common wisdom according to which the properties of PBE water would match those of real water at a simulation temperature $\gtrsim 100$ K above the experimental one is likely too simplistic: PBE water not only freezes at too high temperature, but its dynamics is way too fast at melting, as confirmed by the too-high self-diffusivity predicted by PBE, with respect to SCAN and experiment, when all the simulations are performed at the same temperature offset from $T_m$ as in experiment. For instance, the self-diffusivity of water predicted by PBE at a temperature $T=430~\mathrm K$, which is $\approx 20$ K higher than the PBE melting temperature, $T_m(\mathrm{PBE})\approx 410~\mathrm K$, is 0.45 \AA$^2$\,ps$^{-1}$ \cite{dtisi}. This is to be compared with a value of 0.19 \AA$^2$\,ps$^{-1}$ predicted by SCAN at 20 K above its own melting temperature (\emph{i.e.} at $330\approx 312+20~\mathrm K$, \cite{Chen10846} and practically the same value measured at $T=20\mathrm ~^\circ C$, 0.2 \AA$^2$\,ps$^{-1}$ \cite{Easteal}). In a model where the dependence of the self-diffusivity on temperature were Arrhenius-like, this behaviour would be consistent with a too small pre-exponential factor predicted by PBE relative to SCAN and experiment. Further insight into the dynamics of the water hydrogen-bond network at melting would deserve further investigation.

\begin{figure}[t!]
    \centering
    \includegraphics[width=7.5cm]{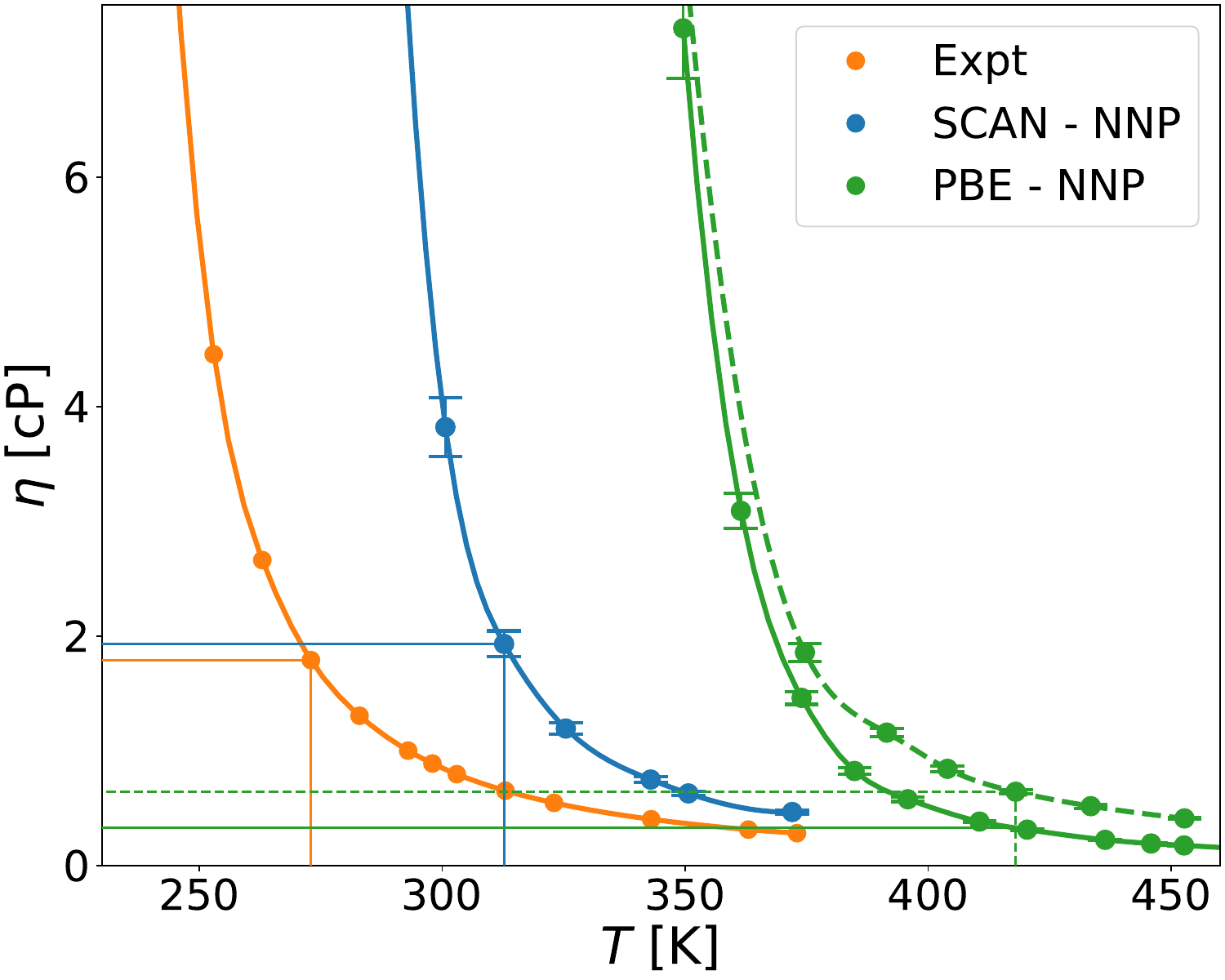}
    \caption{\textbf{Comparison between the shear viscosities of water computed via DPMD simulations using NNP force fields trained to different DFT datasets and with experiments \cite{haynes:2005,Dehaoui12020}.} When not visible, the error bars are smaller than the dots. Continuous lines refer to simulations performed at the equilibrium density corresponding to each temperature. PBE data marked with a dashed line are obtained at the density of 1 gr\,cm$^{-3}$. The thin vertical and horizontal lines mark the melting temperature and the corresponding viscosities. Error bars represent standard deviations.}
    \label{fig:SCANvsPBEvsEXP}
\end{figure}

\begin{figure}[t]
    \centering
    \includegraphics[width=7.5cm]{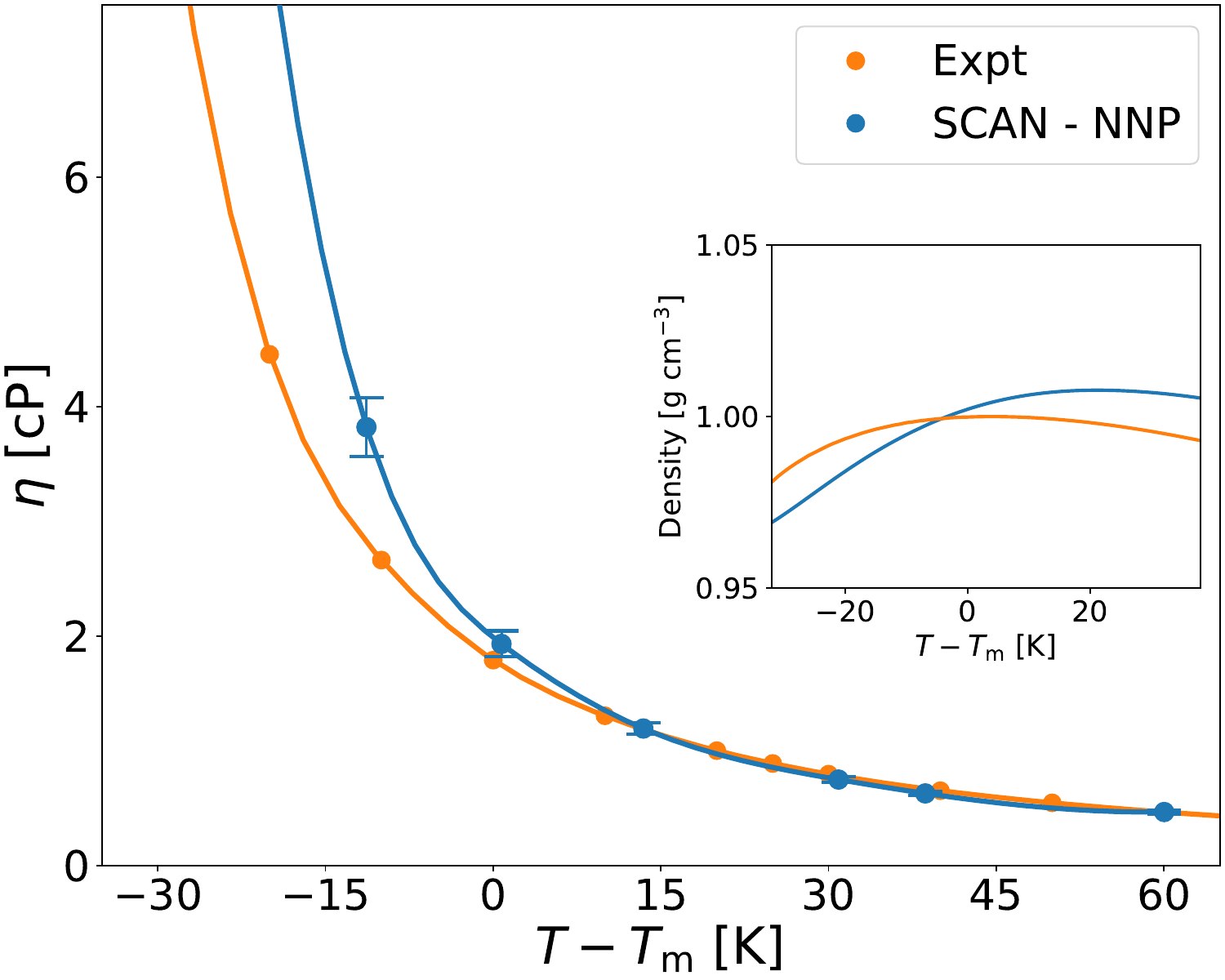}
    \caption{\textbf{Comparison between the SCAN predictions and the experimental values for the shear viscosity of water as a function of the temperature.} The temperature scale for SCAN data has been offset by the difference between the theoretical and experimental melting temperatures, $T_m$ (see text). Error bars represent standard deviations.}
    \label{fig:SCANvsEXP}
\end{figure}

In Fig. 9 we compare with experiment the SCAN-NNP predictions for the viscosity of water, on a temperature scale that has been offset by the difference between the predicted melting temperature for the model and the one observed in experiment, $\Delta T=312-273=39 ~ \mathrm K$. One observes that, while the agreement between theory and experiment is excellent above the melting temperature, SCAN consistently overestimates the viscosity in the undercooled regime. This indicates that the tendency toward dynamical arrest upon undercooling is occurring faster in the model than in experiment. Interestingly, a crossover between the predicted and observed densities occurs at temperatures near melting: SCAN slightly overestimates the density of water for $T>T_m$, while it underestimates it in the undercooled regime. We hypothesize that the too large SCAN predictions for the viscosity below freezing may be related to a propensity of SCAN to overestimate the strength of the hydrogen bonds. In turn, this would lead to overestimate low-density (LD) over high-density (HD) fluctuations upon cooling, corresponding to configurations that underlie the structure of amorphous ices and water. At very deep undercooling they may lead to phase separation between an LD and a HD liquid \cite{sciortino1992,Gartner26040,Car2015}. The stronger local structure of LD water with respect to HD water seems compatible with a more marked solid-like behavior \cite{KUO2021117269,Sciortino2021,Ricci1999} and, hence, with a larger viscosity.

\section{Discussion}\label{sec:conclusion}
We conclude with a summary of our results and some interesting perspective and further applications of our work. In this Article, we have performed a systematic \emph{ab initio} study of the viscosity of liquid water, made possible by a combination of quantum-mechanical first-principles and deep-neural-network techniques. Our study confirms the ability of the SCAN exchange-correlation density functional to predict a broad array of properties of water over a wide range of thermodynamic conditions. Minor shortcomings observed in the undercooled regime are possibly related to the subtle balance between the high- and low-density fluctuations that become more prominent upon undercooling, as one approaches the hypothesized metastable liquid-liquid critical point. These shortcomings might be attenuated by training a neural-network on more accurate quantum mechanical data, such as obtained from hybrid functionals \cite{SCAN0, Gillan2016}, or by using density-corrected DFT (DC-DFT) \cite{Kim2014}, which adopts a more accurate electron density obtained at the Hartree-Fock level of theory.  One of the most successful in describing the property of water is the recent developed DC-SCAN \cite{Paesani}, which produces a remarkably accurate molecular dynamics for liquid water, and a highly realistic self-diffusion coefficient as a function of temperature. Finally as a technical, but important, side product of our study we have highlighted that a careful analysis of the statistical properties of the stress time series, from which the viscosity can be evaluated through the Green-Kubo theory of linear response, is necessary, and we have provided a detailed report on some mathematical and computational tools that can be deployed to ease this task.

\section{\label{sec:II}{Methods}}
The GK theory of linear response \cite{Green,Kubo} provides a rigorous and elegant framework to compute transport coefficients in extended systems, such as the viscosity $\eta$, in terms of the stationary time series of a macroscopic flux \cite{flux} evaluated at thermal equilibrium with MD. For an isotropic system of $N$ interacting particles, the shear viscosity $\eta$ is related to the fluctuations of the off-diagonal elements of the stress tensor:
\begin{equation}
    \label{eq:GK}
        \eta = \frac{V}{k_\mathrm{B} T}\int_0^\infty \langle \, \sigma_{s} \left(\bm{\Gamma}_t \right) \sigma_{s}\left(\bm{\Gamma}_0 \right) \, \rangle \, dt,
\end{equation}
where $V$ is the volume of the system, $T$ is its temperature, $k_\mathrm{B}$ is the Boltzmann’s constant, $\sigma_{s}$ is any of the three independent off-diagonal elements of the stress tensor, ($\sigma_s \in \left\{ \sigma_{xy}, \sigma_{xz}, \sigma_{yz} \right\}$), and $\bm{\Gamma}_t$ indicates the time evolution of a point in phase space from the initial condition $\bm{\Gamma}_0$. In practice, the value of the integral in Equation \eqref{eq:GK} is averaged over the three pairs of Cartesian indices.

\subsection{Expression of stress tensor}
The thermodynamic stress tensor is the equilibrium average, $\bar\sigma$, of a microscopic estimator, $\sigma$, defined as:
\begin{equation}
    \label{eq:stress-def}
    \sigma_{\alpha \beta} = \frac{1}{V}\Bigl[\sum\limits_{i=1}^{N}\frac{p_{i\alpha} p_{i\beta} }{m_i}+\Xi_{\alpha\beta}\Bigr],
\end{equation}
where $\alpha$, $\beta$ represent Cartesian coordinates, $p_{i\alpha}$ is the $\alpha$ component of the momentum of the $i$-th atom, $m_i$ is its mass, while $\Xi$ is the virial term, defined as the derivative of the system's potential energy, $E$, with respect to an uniform scale transformation of the system ($r_\alpha \to r_\alpha +\sum_\beta \epsilon_{\alpha\beta}r_\beta$, $\epsilon$ being the strain tensor):
\begin{equation}
    \Xi_{\alpha \beta} = - \frac{1}{V}\frac{\partial E}{\partial \epsilon_{\alpha \beta} }. \label{eq:virial}
\end{equation}

The expression of the virial term depends on the approach one adopts to perform the simulations: explicit formulas in the classical case are given, \emph{e.g.}, in Ref. \cite{Tsai1979}, for pair-wise potentials, and in Ref. \cite{Fan2015}, for general many-body potentials, while the quantum-mechanical case is thoroughly covered within DFT in Refs. \cite{martin:1983,martin:1985}. The expression of the virial stress using \emph{Deep Potential} models relies on the decomposition of the total energy into individual atomic contributions, as it is the case for the heat current \cite{Fan2015}, and will be presented in some detail in \cref{sec:deepMD}.

\subsection{\label{sec:IIa}{Data Analysis}}
The MD evaluation of the GK formula starts with the computation of the stress time auto-correlation function. This can be done by exploiting the ergodic hypothesis and turning the ensemble average into a time average. The following step is to integrate the tACF, as stated in Equation \eqref{eq:GK}. Despite the apparent simplicity of this process, the straight evaluation of any transport coefficient through the GK formula is jeopardized by the fact that, while ideally the tACF goes to zero for large times, in practice it is very noisy. Indeed, as the tACF approaches zero, Equation \eqref{eq:GK} starts accumulating noise and the integral behaves like the distance traveled by a random walk, whose variance grows linearly with the upper integration limit, making it very difficult to estimate both the bias due to the truncation of the integral and the statistical error.

A better approach is to focus on the \emph{power spectrum} $S(\omega)$ of the stress time series $\sigma_{s}(t)$, which, according to the Wiener-Khintchine theorem \cite{Wiener1930,Khintchine1934}, is the Fourier transform of the tACF of time series:
\begin{equation}
    \label{eq:wiener}
    \begin{aligned}
        S(\omega) & \doteq \lim_{t\to\infty} \frac{1}{t}\left \langle \left | \int_0^t \sigma_s(t') e^{\mathrm{i}\omega t'} dt' \right |^2 \right \rangle \\
        &= \int_{-\infty}^{+\infty}  \langle \sigma_{s}\left(t\right)\sigma_{s}\left(0\right) \rangle e^{\mathrm{i}\omega t} dt.
    \end{aligned}
\end{equation}
According to Equation \eqref{eq:wiener}, the shear viscosity we are after, Equation \eqref{eq:GK}, is proportional to the $\omega=0$ value of the stress power spectrum,
\begin{equation}
    \label{eq:vis-ps}
    \eta = \frac{V}{2 k_\mathrm{B} T}S\left(0\right),
\end{equation}
and any method able to accurately estimate the latter can be leveraged for the former. \emph{Cepstral analysis} \cite{Bogert1963} is one such method \cite{Ercole2017,Baroni2018,Grasselli2021}, and we will rely on it in the present case, as previously done for the thermal and electrical conductivities \cite{Baroni2018,Ercole2017,Bertossa2019,Grasselli2021,Grasselli2020}. A full and user-friendly implementation of cepstral analysis for the estimate of transport coefficients is available in the \sportran\ \cite{SporTran} open-source code.

\subsection{Neural-Network potentials} \label{sec:deepMD}
The \emph{Deep Potential} scheme has already been fully explained in the literature \cite{Linfeng2018,NIPS2018_7696}, so in this section we limit ourselves to a brief overview of its main features.

Let us consider a system of $N$ atoms and let us indicate by $\bm{\mathrm{ R}}$ the set of its atomic coordinates: $\bm{\mathrm{ R}} =\{\bm{\mathrm{ r}}_1,...,\bm{\mathrm{ r}}_N\} \in \mathbb{R}^{3N}$. The potential energy surface of the system $E(\bm{\mathrm{ R}}) =E\left(\bm{\mathrm{ r}}_1,...,\bm{\mathrm{ r}}_N\right)$ is a function of the $3N$ atomic coordinates and of the species of each atom. Assuming that interatomic interactions are local, we make the ansatz that $E(\bm{\mathrm{ R}})$ can be decomposed into the sum of atomic contributions, $\mathcal E_i$, which only depend on the coordinates of the atoms that are close enough to the one they are associated with. In order to establish a convenient notation, let us define by $\bm{\mathcal R}_i$ the set of coordinates of the atoms whose distance from the $i$-th atom is smaller that a certain cut-off radius, $R_c$, referred to the position of the $i$-th atom itself (let $N_i$ be the number of them):
\begin{align}
    \bm{\mathcal R}_i=
    \begin{pmatrix}
\bm{\mathrm{ r}}_{1i} \\
\bm{\mathrm{ r}}_{2i}\\
\vdots \\
\bm{\mathrm{ r}}_{N_i i}\\
\end{pmatrix}=
\begin{pmatrix}
x_{1i} & y_{1i} & z_{1i}\\
x_{2i} & y_{2i} & z_{2i}\\
\vdots & \vdots & \vdots\\
x_{N_i i} & y_{N_i i} & z_{N_i i}\\
\end{pmatrix} \in \mathbb R^{3N_i},
\end{align}
where $\bm{\mathrm{ r}}_{ij}=\bm{\mathrm{ r}}_i-\bm{\mathrm{ r}}_j=\left(x_{ij},y_{ij},z_{ij}\right) $. Using these ingredients, the symmetry-preserving descriptors of the local atomic environments, $\mathcal D_i$, are defined and fed to a neural network, which returns the local atomic energies $\mathcal{E}_{s(i)}(\mathcal D_i)$, depending on the chemical species of the $i$-th atom, $s(i)$, and on its environment, as described by $\mathcal D_i$ (extensive details in Ref. \cite{NIPS2018_7696}). The total potential energy of the system is recovered as the sum of all the atomic contributions, thus ensuring extensivity:
\begin{equation}
    \label{eq:local-contribution}
    E\left(\bm{\mathrm{ R}}\right) = \sum_i \mathcal E_{s(i)}\left({\mathcal{D}}_i\right).
\end{equation}

The neural network is trained to return the local energy contribution corresponding to any given local environment. The training is performed by minimizing the so-called \emph{loss function}, $\mathcal{L}$ with respect to the parameters $\omega$ of the deep-neural network:
\begin{equation}
\label{eq:loss-function}
    \mathcal{L} =p_E \Delta E^2 + \frac{p_F}{3N} \sum_i \Delta\bm{\mathrm{ F}}_i^2, 
\end{equation}
where $\Delta E^2$ and $\Delta \bm{\mathrm{ F}}_i^2$ are the squared deviations of the potential energy and atomic forces respectively, between the reference DFT model and the NNP predictions. The two prefactors, $p_E$ and $p_F$ are needed to optimize the training efficiency and to account for the difference in the physical dimensions of energies and forces.

The force acting on the $i$-th atom is given by:
\begin{equation}
    \begin{aligned}
        \bm{\mathrm{ F}}_i = - \frac{\partial E}{\partial \bm{\mathrm{ r}}_i} & = - \frac{\partial}{\partial \bm{\mathrm{ r}}_i} \sum_j \mathcal{E}_{s(j)}\left(\mathcal{D}_j \right) \\ 
        & =  - \sum_{j} \frac{\partial \mathcal{E}_{s(j)}}{\partial \mathcal{D}_j}\frac{\partial \mathcal{D}_j}{\partial \bm{\mathrm{ r}}_i}
    \end{aligned}
\end{equation}
where we applied Equation \eqref{eq:local-contribution} and the chain rule. Thus the computation of the atomic forces can be split in two different contributions: the first is the derivative of the atomic energy $\mathcal{E}_{s(j)}$ with respect to each element $\mathcal{D}_j$ of the descriptor and can be easily evaluated through TensorFlow \cite{tensorflow2015-whitepaper}, while the second term is given by the gradient of the descriptor with respect to the position of the atoms \cite{Wang2017}.

Beside energies and forces, the NNP predicts also the virial of the system defined as in Equation \eqref{eq:virial}. Using Equation \eqref{eq:local-contribution} one can write\cite{Wang2017}:
\begin{equation}
    \Xi_{\alpha \beta} = \sum_i r_{i\alpha} F_{i\beta} = - \sum_{i \neq j} r_{ij \alpha} \frac{\partial \mathcal{E}_{s(i)}}{\partial r_{ij\beta}},
\end{equation}
where the second term can be further split in two contributions as previously shown for the forces. We remark that the resulting formula is well-defined in PBC and enters directly in the calculation of the stress given by Equation \eqref{eq:stress-def}, serving our purpose of computing the shear viscosity through the GK formula Equation \eqref{eq:GK}.

\section*{Data availability}
Numerical data supporting the plots and relevant results within this paper are available on the Materials Cloud Platform \cite{MatCloud2020,matcloud_cm}. In particular the folder contains: the data training set and the input files for training the PBE-NNP model, and also the \ai stress time series.

\section*{Acknowledgements }
    CM, SB and DT are grateful to Federico Grasselli, Paolo Pegolo and Riccardo Bertossa for enlightening discussions throughout the completion of this work. This work was partially funded by the EU through the \textsc{MaX} Centre of Excellence for supercomputing applications (Project No. 824143) and the Italian MUR, through the PRIN grant \emph{FERMAT}. The work at Princeton University was supported by the Computational Chemical Sciences Center ``Chemistry in Solution and at Interfaces" funded by the US Department of Energy under Award No. DE-SC0019394.

\section*{Author contributions}
Computer simulations and data analysis were mainly performed by CM under the supervision of DT. SB conceived the work and directed some of the data analysis. RC and LZ supervised the early stages of the machine-learning work and provided training data for the SCAN NNP. All authors contributed equally to writing the paper.

\section*{Competing Interests}
No conflict of interests is declared by any of the authors.
\bibliography{biblio}

\end{document}